\documentclass[12pt]{article}
\usepackage{amssymb}
\begin{document}

\def\kbar{{\mathchar'26\mkern-9muk}}  
\def\vev#1{\langle #1 \rangle}
\def\tr{\mbox{Tr}}
\def\ad{\mbox{ad}\,}
\def\ker{\mbox{Ker}\,}
\def\m@th{\mathsurround=0pt}
\def\eqalign#1{\null\,\vcenter{\openup 3pt \m@th
\ialign{\strut\hfil$\displaystyle{##}$&$\displaystyle{{}##}$\hfil
\crcr#1\crcr}}\,}
\def\Dirac{{\raise0.09em\hbox{/}}\kern-0.69em D}

\title{Gravity on Fuzzy Space-Time}

\author{J. Madore \\
        Laboratoire de Physique Th\'eorique et Hautes
Energies\thanks{Laboratoire associ\'e au CNRS, \mbox{URA D0063}} \\
Universit\'e de Paris-Sud, B\^at. 211, F-91405 Orsay \\ }

\date{June, 1997}

\maketitle

\centerline{\it Dedicated to Walter Thirring on the occasion of his 70th
birthday}

\vskip 2cm

\abstract{A review is made of recent efforts to add a gravitational
field to noncommutative models of space-time.  Special emphasis is
placed on the case which could be considered as the noncommutative
analog of a parallelizable space-time. It is argued that, at least in
this case, there is a rigid relation between the noncommutative
structure of the space-time on the one hand and the nature of the
gravitational field which remains as a `shadow' in the commutative limit
on the other.}

\vfill
\noindent
ESI Preprint 478 (1997). Lecture given at the International Workshop
``Mathematical Physics - today, Priority Technologies - for tomorrow'',
Kyiv, Ukraine, May 1997.
\medskip
\eject

\parskip 4pt plus2pt minus2pt

\section{Introduction and Motivation}

Simply stated, `fuzzy space-time' is a space-time in which the
`coordinates' do not commute. One typically replaces the four Minkowski
coordinates $x^\mu$ by four generators $q^\mu$ of a noncommutative
algebra which satisfy commutation relations of the form 
$$ 
[q^\mu, q^\nu] = i \kbar q^{\mu\nu}.                               \eqno(1.1) 
$$ 
The parameter $\kbar$ is a fundamental area scale which we shall suppose
to be of the order of the Planck area:
$$
\kbar \simeq \mu_P^{-2} = G\hbar.
$$ 
Equation~(1.1) contains in fact little information about the algebra.
If the right-hand side does not vanish it states that at least some of
the $q^\mu$ do not commute. It states also that it is possible to
identify the original coordinates with the generators $q^\mu$ in the
limit where the Planck mass $\mu_P$ tends to infinity: 
$$ 
x^\mu = \lim_{\kbar \rightarrow 0} q^\mu.                          \eqno(1.2) 
$$ 
For mathematical simplicity we shall suppose this to be the case
although one could include a singular `renormalization constant' $Z$ and
replace (1.2) by an equation of the form
$$ 
Z \, x^\mu = \lim_{\kbar \rightarrow 0} q^\mu.                      \eqno(1.3) 
$$ 
If, as we shall argue, gravity acts as a universal regulator for
ultraviolet divergences then one could reasonably expect the limit
$\kbar \rightarrow 0$ to be a singular limit. An argument in this sense
has been given by Mangano~\cite{Man97}.

Perhaps not the simplest but certainly the most familiar example of a
`fuzzy space' is the quantized version of a 2-dimensional phase space,
described by the `coordinates' $p$ and $q$. This example has the
advantage of illustrating what is for us the essential interest of the
relation of the form (1.1) as expressed in the Heisenberg uncertainty
relations. Since one cannot measure simultaneously $p$ and $q$ to
arbitrary precision quantum phase space has no longer a notion of a
point.  It can however be thought of as divided into cells of volume
$2\pi\hbar$. If the classical phase space is of finite total volume
there will be a finite number of cells and the quantum system will have
a finite number of possible states. A `function' then on quantum phase
space will be defined by a finite number of values and can be
represented by a matrix.

By analogy with quantum mechanics we shall suppose that the generators
$q^\mu$ can be represented as hermitian operators on some (complex)
Hilbert space.  The presence of the factor $i$ in (1.1) implies that the
$q^{\mu\nu}$ are also hermitian operators.  The $q^\mu$ have real
eigenvalues but because of the relations (1.1) they cannot be
simultaneously diagonalized; points are ill-defined and space-time
consists of elementary cells of volume $(2 \pi \kbar)^2$.  Now when a
physicist calculates a Feynman diagram he is forced to place a cut-off
$\Lambda$ on the momentum variables in the integrands.  This means that
he renounces any interest in regions of space-time of volume less than
$\Lambda^{-4}$.  As $\Lambda$ becomes larger and larger the forbidden
region becomes smaller and smaller but it can never be made to vanish.
There is a fundamental length scale, much larger than the Planck length,
below which the notion of a point is of no practical importance.  The
simplest and most elegant, if certainly not the only, way of introducing
such a scale in a Lorentz-invariant way is through the introduction of
the `coordinates' $q^\mu$. The analogs of the Heisenberg uncertainty
relations imply then that 
$$
\Lambda^2 \kbar \lesssim 1.
$$ 
The existence of a forbidden region around each point in space-time
means that the standard description of Minkowski space as a
4-dimensional continuum is redundant. There are too many points.
Heisenberg already in the early days of quantum field theory proposed to
replace the continuum by a lattice structure.  A lattice however breaks
Poincar\'e invariance and can hardly be considered as fundamental.  It
was Snyder~\cite{Sny47a} who first had the idea of using non-commuting
coordinates to mimic a discrete structure in a covariant way although
something similar had previously been proposed by Markov~\cite{Mar40}.
In his article~\cite{Fin69} on the subject Finkelstein cites Riemann as
the first person to be concerned with the existence of discrete objects
within a continuum space.

As a simple illustration of how a `space' can be `discrete' in some
sense and still covariant under the action of a continuous symmetry
group one can consider the ordinary round 2-sphere, which has acting on
it the rotational group $SO_3$. As a simple example of a lattice
structure one can consider two points on the sphere, for example the
north and south poles.  One immediately notices of course that by
choosing the two points one has broken the rotational invariance. It can
be restored at the expense of commutativity. The set of functions on the
two points can be identified with the algebra of diagonal $2 \times 2$
matrices, each of the two entries on the diagonal corresponding to a
possible value of a function at one of the two points. Now an action of
a group on the lattice is equivalent to an action of the group on the
matrices and there can obviously be no non-trivial action of the group
$SO_3$ on the algebra of diagonal $2 \times 2$ matrices. However if one
extends the algebra to the noncommutative algebra of all $2 \times 2$
matrices one recovers the invariance. The two points, so to speak, have
been smeared out over the surface of a sphere; they are replaced by two
cells.  An `observable' is an hermitian $2 \times 2$ matrix and has
therefore two real eigenvalues, which are its values on the two cells.
Although what we have just done has nothing to do with Planck's constant
it is similar to the procedure of replacing a classical spin which can
take two values by a quantum spin of total spin 1/2. Only the latter is
invariant under the rotation group. By replacing the spin 1/2 by
arbitrary spin $s$ one can describe a `lattice structure' of $n = 2s+1$
points in an $SO_3$-invariant manner. The algebra becomes then the
algebra $M_n$ of $n \times n$ complex matrices. We shall discuss this
example in more detail in Section~5.3.

It is to be stressed that we modify the structure of Minkowski
space-time but maintain covariance under the action of the Poincar\'e
group. A fuzzy space-time looks then like a solid which has a
homogeneous distribution of dislocations but no disclinations. We can
pursue this solid-state analogy and think of the ordinary Minkowski
coordinates as macroscopic order parameters obtained by coarse-graining
over scales less than the fundamental scale.  They break down and must
be replaced by elements of some noncommutative algebra when one
considers phenomena on these scales. It might be argued that since we
have made space-time `noncommutative' we ought to do the same with the
Poincar\'e group. This logic leads naturally to the notion of a
$q$-deformed Poincar\'e (or Lorentz) group which act on a very
particular noncommutative version of Minkowski space called
$q$-Minkowski space. We discuss $q$-deformations in Section~5.2. It has
also been argued, for conceptual as well as practical, numerical
reasons, that the lattice version of space-time or of space is quite
satisfactory if one uses a random lattice structure or graph. From this
point of view the Lorentz group is a classical invariance group and is
not valid at the microscopic level. We shall briefly mention this
possibility in Section~5.1.

Let ${\cal A}_\kbar$ be the algebra generated in some sense by the
elements $q^\mu$. We shall be here working on a formal level so that one
can think of ${\cal A}_\kbar$ as an algebra of polynomials in the
$q^\mu$ although we shall explicitly suppose that there are enough
elements to generate smooth functions on space-time in the commutative
limit.  Since we have identified the generators as hermitian operators
on some Hilbert space we can identify ${\cal A}_\kbar$ as a subalgebra
of the algebra of all operators on the Hilbert space. We have added the
subscript $\kbar$ to underline the dependence on this parameter but of
course the commutation relations (1.1) do not determine the structure of
${\cal A}_\kbar$, We in fact conjecture that every possible
gravitational field can be considered as the commutative limit of a
noncommutative equivalent and that the latter is strongly restricted if
not determined by the structure of the algebra ${\cal A}_\kbar$. We must
have then a large number of algebras ${\cal A}_\kbar$ for each value of
$\kbar$.

We argued above that the noncommutative structure gives rise to an
ultraviolet cut-off. This idea has been developed by several
authors~\cite{HelTan54, DopFreRob95, KemManMan95, KemMan96} since the
original work of Snyder~\cite{Sny47a,Sny47b}.  It is the right-hand
arrow of the diagram 
$$
\def\normalbaselines{\baselineskip=18pt}
\matrix{
{\cal A}_\kbar &\Longleftarrow &\Omega^*({\cal A}_\kbar)\cr
\Downarrow && \Uparrow \cr
\hbox{Cut-off} &&\hbox{Gravity}
}
\def\normalbaselines{\baselineskip=12pt}                            \eqno(1.4)
$$ 
The top arrow is a mathematical triviality; the $\Omega^*({\cal
A}_\kbar)$ is what gives a differential structure to the algebra. We
shall define and discuss it in Section~2.2.  The main section is
Section~4. In it we shall attempt, not completely successfully, to argue
that each gravitational field is the unique `shadow' in the limit $\kbar
\rightarrow 0$ of some differential structure over some noncommutative
algebra. This would define the left-hand arrow of the diagram.

The composition of the three arrows is an expression of an old idea, due
to Pauli and developed by Deser~\cite{Des57} and
others~\cite{IshSalStr71}, that perturbative ultraviolet divergences
will one day be regularized by the gravitational field. The possibility
which we shall consider here is that the mechanism by which this works
is through the introduction of noncommuting `coordinates' such as the
$q^\mu$. A hand-waving argument can be given~\cite{MadMou95} which
allows one to think of the noncommutative structure of space-time as
being due to quantum fluctuations of the light-cone in ordinary
4-dimensional space-time.  This relies on the existence of quantum
gravitational fluctuations. A purely classical argument based on the
formation of black-holes has been also given~\cite{DopFreRob95}.  In
both cases the classical gravitational field is to be considered as
regularizing the ultraviolet divergences through the introduction of the
noncommutative structure of space-time. This can be strengthened as the
conjecture that the classical gravitational field and the noncommutative
nature of space-time are two aspects of the same thing. It is our
purpose here to explore in some detail this relation.

For an sampling of the early history of ideas on the microtexture of
space-time we refer to Section~1.3 of the book by Prugove\v
cki~\cite{Pru95} as well as to the review articles by Kragh \&
Carazza~\cite{KraCar94} and Gibbs~\cite{Gib95}.  When referring to the
version of space-time which we describe here we use the adjective
`fuzzy' to underline the fact that points are ill-defined.  Since the
algebraic structure is described by commutation relations the qualifier
`quantum' has also been used~\cite{Sny47a, DopFreRob95, MadMou96b}. This
latter expression is unfortunate since the structure has no immediate
relation to quantum mechanics and also it leads to confusion with
`spaces' on which `quantum groups' act.  To add to the confusion the
word `quantum' has also been used~\cite{GreYau97} to designate
equivalence classes of ordinary differential geometries which yield
isomorphic string theories and the word `lattice' has been
used~\cite{'tH96} to designate what we here qualify as `fuzzy'. The idea
of a $q$-deformation goes back to the dawn of time.  Almost immediately
after Clifford introduced his algebras they were $q$-deformed with $q$ a
root of unity by Sylvester~\cite{Syl84} and by Cartan~\cite{Car98}. This
idea was taken up later in a special case by Weyl~\cite{Wey50} and
Schwinger~\cite{Sch60} to produce a finite version of quantum mechanics.

\section{Fuzzy space-time}

\subsection{Space-time as an algebraic structure}

We saw in the Introduction that by making the coordinates noncommutative
we lost the space-time but retained an equivalent of the algebra of
functions on it.  The purpose of noncommutative geometry is to
reformulate as much as possible the geometry of a space in terms of its
algebra of functions and then generalize the corresponding results of
differential geometry to the case of a noncommutative algebra. We have
noticed that the main notion which is lost when passing from the
commutative to the noncommutative case is that of a point.
`Noncommutative geometry is pointless geometry.'  The original
noncommutative geometry is based on the quantized phase space of
non-relativistic quantum mechanics.  In fact Dirac in his historical
papers in 1926~\cite{Dir26a, Dir26b} was aware of the possibility of
describing phase-space physics in terms of the quantum analog of the
algebra of functions, which he called the quantum algebra and he was
aware of the absence of localization, expressed by the Heisenberg
uncertainty relation, as a central feature of these geometries. Inspired
by work by von Neumann, for several decades physicists studied quantum
mechanics and quantum field theory as well as classical and quantum
statistical physics giving prime importance to the algebra of
observables and considering the state vector as a secondary derived
object. This work has much in common with noncommutative geometry. The
notion of a pure state replaces that of a point.

The details of the structure of the algebra ${\cal A}_\kbar$ will be
contained, for example, in the commutation relations $[q^\lambda,
q^{\mu\nu}]$.  The $q^{\mu\nu}$ can be also considered as extra
generators and the Equations~(1.1) as extra relations. In this case the
$q^{\mu\nu}$ cannot be chosen arbitrarily. They must satisfy the four
Jacobi identities: 
$$ [q^\lambda, q^{\mu\nu}] + [q^\mu, q^{\nu\lambda}]
+ [q^\nu, q^{\lambda\mu}] = 0.                                     \eqno(2.1) 
$$ 
One can define recursively
an infinite sequence of elements by setting, for $p \geq 1$, 
$$
[q^\lambda, q^{\mu_1 \cdots \mu_p}] = i \kbar q^{\mu_1 \cdots
\mu_{(p+1)}}.                                                      \eqno(2.2) 
$$ 
Several structures have been considered in
the past~\cite{Sny47a, Mad89a, DopFreRob95, Mad95}. With our choice of
normalization $q^{\mu_1 \cdots
\mu_p}$ has units of mass to the power $p-2$. We shall assume that for 
the description of a generic gravitational field the appropriate algebra
${\cal A}_\kbar$ has a trivial center, that the only elements which
commute with all other elements are the constant multiples of the
identity element.  The only argument we have in favour of this
assumption is the fact that it could be argued that if the center is not
trivial then the `quantization' has been only partial.  It implies of
course that the sequence of $q^{\mu_1 \cdots \mu_p}$ never ends,
although all these elements need not be independent.  The observables
will be some subset of the hermitian elements of ${\cal A}_\kbar$.  We
shall not discuss this problem here; we shall implicitly suppose that
all hermitian elements of ${\cal A}_\kbar$ are observables, including
the `coordinates'.  We shall not however have occasion to use explicitly
this fact.

Consider the structure of the `classical' limit ${\cal A}_0$ of ${\cal
A}_\kbar$ obtained by letting $\kbar \rightarrow 0$. For this we must
suppose that $Z = 1$ in Equation~(1.3).  Assume that one can identify
${\cal A}_0 = {\cal C}(V_0)$ as the algebra of smooth, complex-valued
functions on a real extension $V_0$ of space-time of dimension $\geq 4$.
and that there is a projection of $V_0$ onto ordinary space-time.  We
set $$ x^{\mu_1 \cdots \mu_p} =
\lim_{\kbar \rightarrow 0}q^{\mu_1 \cdots \mu_p}.
$$ A set of independent elements of the complete set of the $x^{\mu_1
\cdots \mu_p}$ are local coordinates of $V_0$. The dimension of $V_0$
will depend on how many there are. If space-time is Minkowski space-time
then the condition of Lorentz invariance in the commutative limit forces
$x^\lambda$ and at least 4 of the 6 coordinates $x^{\mu\nu}$ to be
independent~\cite{DopFreRob95, DubKerMad97}. In general the set
$x^{\mu_1 \cdots \mu_p}$ for $p \geq 3$ can at least in part be
functions of $x^\lambda$ and $x^{\mu\nu}$.  One can consider $V_0$ as a
Kaluza-Klein extension of space-time by a space which is perhaps of
infinite dimension and in general not compact.  It should be stressed
however that $V_0$ is a mathematical fiction. The `real' world is
described by the algebra ${\cal A}_\kbar$; it is this algebra which we
consider to be the correct Kaluza-Klein extension of
space-time~\cite{Mad89b, MadMou95}.  The difference in dimension between
$V_0$ and space-time is one of the measures of the extent to which the
verb `to quantize' as applied to the coordinates of space-time is a
misnomer; one could {\it in extremis} `quantize' the coordinates of
$V_0$. Even here the verb should be restricted to cases in which the
right-hand side of (1.1) lies in the center of the algebra.

Quite generally the commutator of ${\cal A}_\kbar$ defines a Poisson
structure on $V_0$.  We have given arguments~\cite{MadMou96b}, based on
simple models, that a differential calculus over ${\cal A}_\kbar$ should
determine a metric-compatible torsion-free linear connection on $V_0$.
It is natural then that there should be a relation between the Poisson
structure and the curvature of the connection. One can
show~\cite{Mad97a} that certain natural hypotheses on the differential
calculus yield in fact relations between the two.  We have been however
unable so far to present a realistic gravitational field explicitly as
the `shadow' of a differential calculus over a noncommutative algebra.

If there is a gravitational field then there must be some source, of
characteristic mass $\mu$. If $\mu^2 \kbar$ tends to zero with $\kbar$
then $V_0$ will be without curvature. This case has been considered
previously~\cite{DopFreRob95, DubKerMad97}. We are interested here in
the case in which $\mu^2 \kbar$ tends to some finite non-vanishing value
as $\kbar \rightarrow 0$.

\subsection{Space-time as a differential structure}

But space-time is more than just an algebra of functions; it has a
differential structure. Although it was von~Neumann who introduced the
expression `noncommutative geometry' it was only recently that
mathematicians, notably Connes~\cite{Con86, Con94}, have developed the
theory of `differential noncommutative geometry' which we shall use
here. The central notion is that of a differential form.  We shall
define a differential by a set of simple rules which makes it obvious
that it is equivalent to a derivative and ask the reader to believe that
the rules have a rigorous and natural mathematical foundation. He will
see that they are quite easy to manipulate in the simple noncommutative
geometries we consider.

We first recall the commutative case. The set of smooth functions
${\cal A}$ on space-time is a commutative algebra, which it is
convenient to consider over the complex numbers.  A 1-form is a
covariant vector field $A_\mu$, which we shall write as $A = A_\mu
dx^\mu$ using a set of basis elements $dx^\mu$. A 2-form is an
antisymmetric 2-index covariant tensor $F_{\mu\nu}$ which we shall write
as
$$
F = {1\over 2} F_{\mu\nu} dx^\mu dx^\nu
$$
using the product of the basis elements. This product is antisymmetric 
$$
dx^\mu dx^\nu = - dx^\nu dx^\mu                                    \eqno(2.3)
$$
but otherwise has no relations.  Higher-order forms can be defined as
arbitrary linear combination of products of 1-forms. A $p$-form can be
thus written as
$$
\alpha = {1\over p!} \alpha_{\lambda_1 \cdots \lambda_p} 
dx^{\lambda_1} \cdots dx^{\lambda_p}.
$$
The coefficients $\alpha_{\lambda_1 \cdots \lambda_p}$ are smooth 
functions and completely antisymmetric in the $p$ indices. 

We define $\Omega^0({\cal A}) = {\cal A}$ and for each $p$ we write the
vector space of $p$-forms as $\Omega^p({\cal A})$. Each 
$\Omega^p({\cal A})$ depends obviously on the algebra ${\cal A}$ and,
what is also obvious and very important, it can be multiplied both from
the left and the right by the elements of ${\cal A}$; each
$\Omega^p({\cal A})$ is an ${\cal A}$-bimodule.  It is easy to see that
$\Omega^p({\cal A}) = 0$ for all $p \geq 5$.  We define 
$\Omega^*({\cal A})$ to be the set of all $\Omega^p({\cal A})$.  It has
a product $\pi$ induced by (2.3); it is a graded commutative algebra.  The
product defines a projection
$$
\Omega^1({\cal A}) \otimes_{\cal A} \Omega^1({\cal A}) 
\buildrel \pi \over \longrightarrow \Omega^2({\cal A}).
$$
The algebra $\Omega^*({\cal A})$ can be written as a sum
$$
\Omega^*({\cal A}) = \Omega^+({\cal A}) \oplus \Omega^-({\cal A})   \eqno(2.4)
$$
of even forms and odd forms. The $A$ is an odd form and $F$ is even.
The algebra ${\cal A}$ is a subalgebra of $\Omega^+({\cal A})$.

Let $f$ be a function, an element of the algebra 
${\cal A} = \Omega^0({\cal A})$.  We define a map $d$ from
$\Omega^p({\cal A})$ into $\Omega^{p+1}({\cal A})$ by the rules
$$
df = \partial_\mu f dx^\mu, \qquad d^2 = 0.                         \eqno(2.5)
$$
It takes odd (even) forms into even (odd) ones. From the rules we
find that
$$
d A = d(A_\mu dx^\mu) 
= {1\over 2}(\partial_\mu A_\nu - \partial_\nu A_\mu) dx^\mu dx^\nu = F
$$
if we set
$$
F_{\mu\nu} = \partial_\mu A_\nu - \partial_\nu A_\mu.
$$
From the second rule we have 
$$
d F = 0.
$$
It is easy to see that if $\xi$ is a $p$-form and $\eta$ is a 
$q$-form then
$$
\xi \eta = (-1)^{pq} \eta \xi, \qquad
d(\xi \eta) = (d \xi) \eta + (-1)^p \xi d\eta.
$$

The couple $(\Omega^*({\cal A}), d)$ is called the de~Rham differential
algebra or the de~Rham differential calculus over ${\cal A}$.  What
distinguishes it is the fact that the 1-forms are dual to the Lie
algebra of derivations of ${\cal A}$. The derivative $\partial_\mu f$ of
a smooth function $f$ is a smooth function. We use the word derivation
to distinguish the map $\partial_\mu$ from the result of the map
$\partial_\mu f$. A general derivation is a linear map $X$ from the
algebra into itself which satisfies the Leibniz rule: 
$X(fg) = (Xf)g + f(Xg)$. In the case we are presently considering a
derivation can always be written in terms of the basis $\partial_\mu$ as
$X = X^\mu \partial_\mu$. Such is not always the case.  The relation
between $d$ and $\partial_\mu$ is given by
$$
df (\partial_\mu) = \partial_\mu f.
$$
This equation has the same content as the first of the relations~(2.5).
One passes from one to the other by using the particular case
$$
dx^\mu (\partial_\nu) = \delta^\mu_\nu.                           \eqno(2.6)
$$
The basis $dx^\mu$ is said to be dual to the basis $\partial_\mu$.  The
derivations form a vector space (the tangent space) at each point, and
(2.6) defines $df$ as an element of the dual vector space (the cotangent
space) at the same point.  Over an arbitrary algebra which has
derivations one can always define in exactly the same manner a
differential calculus based on derivations.  These algebras have thus at
least two, quite different, differential calculi, the universal one and
the one based on the set of all derivations.

Over each algebra ${\cal A}$, be it commutative or not, there can exist
a multitude of differential calculi.  This fact makes the noncommutative
version of geometry richer than the commutative version. As a simple
example we define what is known as the universal calculus
$(\Omega_u^*({\cal A}), d_u)$ over the commutative algebra of functions
${\cal A}$. We set, as always, $\Omega_u^0({\cal A}) = {\cal A}$ and for
each $p \geq 1$ we define $\Omega_u^p({\cal A})$ to be the set of
$(p+1)$-point functions which vanish when any two points coincide. It is
obvious that $\Omega_u^p({\cal A}) \neq 0$ for all $p$.  There is a map
$d_u$ from $\Omega_u^p({\cal A})$ into $\Omega_u^{p+1}({\cal A})$ given
by $(d_uf)(x^\mu, y^\mu) = f(y^\mu) - f(x^\mu)$ for $p = 0$.  This can
also be written without reference to points as
$$
d_uf = 1 \otimes f - f \otimes 1.
$$
For $p \geq 1$, $d_u$ is given by a similar sort of alternating sum
defined so that $d_u^2 = 0$.  The algebra $\Omega_u^*({\cal A})$ is not
graded commutative.  It is defined for arbitrary functions, not
necessarily smooth, and it has a straightforward generalization to
arbitrary algebras, not necessarily commutative.

To explain the qualifier `universal' let $(\Omega^*({\cal A}), d)$ be
any other differential calculus over ${\cal A}$, for example the usual
de~Rham differential calculus. Then there is a unique $d_u$-homomorphism
$\phi$
$$
\Omega^*_u({\cal A}) \buildrel \phi \over \longrightarrow  \Omega^*({\cal A})  
$$
of $\Omega^*_u({\cal A})$ onto $\Omega^*({\cal A})$. It is given by
$$
\phi (f) =  f,  \qquad \phi (d_u f) =  d f.
$$
If we choose a coordinate system and expand the function $f(y^\mu)$ about 
the point $x^\mu$,
$$
f(y^\mu) = f(x^\mu) + (y^\nu - x^\nu) \partial_\nu f(x^\mu) + \cdots,
$$
we see that the map $\phi$ is given by
$$
\phi (y^\mu - x^\mu) = dx^\mu
$$
and that it annihilates any 1-form $f(x^\mu,y^\mu) \in \Omega^1_u({\cal A})$
which is second order in $y^\mu- x^\mu$. One such form is $fd_ug - d_ugf$,
given by
$$
(fd_ug -d_ugf)(x^\mu, y^\mu) = 
- (f(y^\mu) - f(x^\mu))(g(y^\mu) - g(x^\mu)).
$$
It does not vanish in $\Omega^1_u({\cal A})$ but its image in 
$\Omega^1({\cal A})$ under $\phi$ is equal to zero.

To form tensors one must be able to define tensor products, for
example the tensor product 
$\Omega^1({\cal A}) \otimes_{\cal A} \Omega^1({\cal A})$ of 
$\Omega^1({\cal A})$ with itself.
We have here written in subscript the algebra ${\cal A}$. This
piece of notation indicates the fact that we identify $\xi f \otimes \eta$ 
with $\xi \otimes f \eta$ for every element $f$ of the algebra, a
technical detail which is important. It means also that one must be
able to multiply the elements of $\Omega^1({\cal A})$ on the left
and on the right by the elements of the algebra ${\cal A}$. Since ${\cal A}$ 
is commutative of course these two operations are equivalent and
this left (right) linearity is equivalent to the property of locality. It
means that the product of a function with a 1-form at a point is again a
1-form at the same point, a property which distinguishes the ordinary
product from other, non-local, products such as the convolution. In the
noncommutative case there are no points and locality can not be defined;
it is replaced by the property of left and right linearity with
respect to the algebra.

There is an interesting relation between the differential $d$ and the
Dirac operator $\Dirac$.  Let $\psi$ be a Dirac spinor and $f$ a smooth
function. It is straightforward to see that
$$
\partial_\lambda f \gamma^\lambda\psi = - [i\Dirac, f] \psi.
$$
If we make the replacement $\gamma^\lambda \mapsto dx^\lambda$ the left-hand
side becomes equal to $df \psi$ and we can write the differential as a
commutator: 
$$
df = - [i\Dirac, f].
$$ 
It would be natural to try to generalize this relation to higher-order
forms by using a graded commutator on the right-hand side.
Because $dx^\mu dx^\nu + dx^\nu dx^\mu = 0$ whereas
$\gamma^\mu \gamma^\nu + \gamma^\nu \gamma^\mu \neq 0$ one would find
that $d^2 \neq 0$. This problem is connected with the fact that the
square of the Dirac operator is not proportional to the identity.
We shall see below a simpler example of this and mention how
one solves the problem.

Consider now the noncommutative algebra ${\cal A}_\kbar$. First we note
that if $X$ is a derivation of ${\cal A}_\kbar$ and $f$ an arbitrary
element then in general $fX$ is no longer a derivation. The simplest
examples to see this are the matrix algebras which we shall mention
below. Also it is sometimes of interest to consider algebras which have
no derivations.  It is for these reasons that derivations do not play
the same role in noncommutative geometry which vector fields play in
ordinary geometry and it is much more convenient to use a differential 
calculus to describe the differential structure.

Suppose that a set of 1-forms $\Omega^1({\cal A}_\kbar)$ has been
constructed and that there is a map $d$ of 
$\Omega^0({\cal A}_\kbar) = {\cal A}_\kbar$ into 
$\Omega^1({\cal A}_\kbar)$:
$$
\Omega^0({\cal A}_\kbar) \buildrel d \over \longrightarrow 
\Omega^1({\cal A}_\kbar).                                          \eqno(2.7)
$$
We shall construct a differential calculus over ${\cal A}_\kbar$ using a
procedure due to Connes and Lott~\cite{ConLot92}.  In the form which we
shall use it the procedure has been described in detail
elsewhere~\cite{DimMad96} but the idea is simple. As in the commutative
case we suppose that the $\Omega^1({\cal A}_\kbar)$ has the structure of
an ${\cal A}_\kbar$-bimodule, that an element of $\Omega^1({\cal A}_\kbar)$ 
can be multiplied from the right and from the left by an
arbitrary element of ${\cal A}_\kbar$ and the result is still an element
of $\Omega^1({\cal A}_\kbar)$.  We define the bimodule of 2-forms to be
the largest set of elements of the form $dfdg$ with a product between
$df$ and $dg$ subject only to the condition that it be consistent with
the bimodule structure of $\Omega^1({\cal A}_\kbar)$ and with the condition 
$d^2 = 0$. We have
then
$$
\Omega^0({\cal A}_\kbar) \buildrel d \over \longrightarrow 
\Omega^1({\cal A}_\kbar) \buildrel d \over \longrightarrow
\Omega^2({\cal A}_\kbar).
$$
If, for example, $f dg - dg f = 0$ as in the commutative case then we
must have $d(f dg - dg f) = df dg + dg df = 0$.  This construction can
be continued to arbitrary $p$-forms~\cite{DimMad96}. It is of course
perfectly consistent to choose a smaller algebra of forms. One could
set, for example, $\Omega^p({\cal A}_\kbar) = 0$ for all $p \geq 2$.

We shall find it convenient to define $\Omega^1({\cal A}_\kbar)$ in
terms of a set of derivations of ${\cal A}_\kbar$.  For each integer $n$
let $\lambda_i$ be a set of $n$ linearly independent antihermitian
elements of ${\cal A}_\kbar$ and introduce the derivations $e_i$ defined
by
$$
e_i f = [\lambda_i, f].
$$
The Leibniz rule follows from the Jacobi identity for the bracket.  In
general the $e_i$ do not form a Lie algebra but they do however satisfy
commutation relations as a consequence of the commutation relations of
${\cal A}$. In order for them to have the correct dimensions one must
introduce a mass parameter $\mu$ and replace $\lambda_i$ by 
$\mu \lambda_i$. We shall set $\mu = 1$.  We shall suppose that if an
element of ${\cal A}_\kbar$ commutes with all of the $\lambda_i$ then it
is a constant multiple of the unit element. This is the noncommutative
equivalent of the statement that a function is a constant if all of its
partial derivatives vanish.  Define $\Omega^1({\cal A}_\kbar)$ and the
map (2.7) by
$$
df (e_i) = e_i \, f.                                               \eqno(2.8)
$$

We shall suppose that there exists a set of $n$ elements $\theta^i$ of
$\Omega^1({\cal A}_\kbar)$ such that
$$
\theta^i (e_j) = \delta^i_j.                                       \eqno(2.9)
$$
In the examples which we consider we shall show that the $\theta^i$
exist by explicit construction.  We shall refer to the set of
$\theta^i$ as a frame or Stehbein.  It commutes with all the elements
$f$ of ${\cal A}_\kbar$:
$$
f \theta^i = \theta^i f.                                           \eqno(2.10)
$$
This follows directly from (2.9) and from the definition of the module
structure:
$$
f dg (e_i) = f e_i \, g, \qquad (dg) f (e_i) =  (e_i \, g) f.
$$
The ${\cal A}$-bimodule $\Omega^1({\cal A})$ is generated by all
elements of the form $f dg$ or of the form $df g$. Because of the
Leibniz rule these conditions are equivalent.  Using the frame we can
write
$$
f dg = (f e_i g) \theta^i, \qquad (dg) f =  (e_i g) f \theta^i.    \eqno(2.11)
$$
The commutation relations of the algebra constrain then the relations
between $f dg$ and $dg f$ for all $f$ and $g$.  

Because of the commutation relations of the algebra the $\theta^i$
satisfy in general commutation relations. Since 
$\Omega^*({\cal A}_\kbar)$ is an algebra there is a natural product
map
$$
\Omega^1({\cal A}_\kbar) \otimes_{{\cal A}_\kbar}
\Omega^1({\cal A}_\kbar) \buildrel \pi \over \longrightarrow
\Omega^2({\cal A}_\kbar)
$$
We shall suppose that $\Omega^2({\cal A}_\kbar)$ is a submodule of
$\Omega^1({\cal A}_\kbar) \otimes_{{\cal A}_\kbar} \Omega^1({\cal A}_\kbar)$ 
and that $\pi$ is a projection. We can write therefore
$$
\pi(\theta^i \otimes \theta^j) = 
P^{ij}{}_{kl} \theta^k \otimes \theta^l                           \eqno(2.12)
$$
where, because of (2.10), the $P^{ij}{}_{kl}$ are complex numbers 
with
$$
P^{ij}{}_{mn} P^{mn}{}_{kl} =  P^{ij}{}_{kl}.                      \eqno(2.13)
$$
The product $\theta^i \theta^j$ satisfies then the relations
$$
\theta^i \theta^j = P^{ij}{}_{kl} \theta^k \theta^l.               \eqno(2.14)
$$
In several important cases which we shall consider the $\theta^i$
anticommute. This corresponds to the expression 
$$
P^{ij}{}_{kl} = 
{1\over 2} (\delta^i_k \delta^j_l - \delta^j_k \delta^i_l)         \eqno(2.15)
$$
for the $P^{ij}{}_{kl}$.

Define $\theta = - \lambda_i \theta^i$. Then one sees that
$$
df = e_i f \theta^i = - [\theta, f]                                \eqno(2.16)
$$
and it follows that as a bimodule $\Omega^1({\cal A})$ is generated by
one element. We see also that $\theta$ plays a role in these
differential calculi that the Dirac operator does in the commutative
case.  The $\theta$ is here however itself an element of 
$\Omega^1({\cal A})$ whereas $i \Dirac$ cannot be considered as a 
1-form.  Under the condition (2.9) the $\Omega^1({\cal A})$ is free of
rank $n$ as a left or right module. It can therefore be identified with
the direct sum of $n$ copies of ${\cal A}_\kbar$:
$$
\Omega^1({\cal A}_\kbar) = \bigoplus_1^n {\cal A}_\kbar.          \eqno(2.17)
$$
This equation states that a 1-form can be described by its components.
It implies that the fuzzy space-times which we consider are the
noncommutative equivalents of parallelizable manifolds.
We see that the rank of $\Omega^1({\cal A})$ can be an arbitrary integer. 

One can show~\cite{MadMou96b} that $\theta$ satisfies the equation
$$
d\theta + \theta^2 = - {1\over 2} K_{ij} \theta^i \theta^j        \eqno(2.18)
$$
where the $K_{ij}$ are complex numbers. One can show
further~\cite{DimMad96, MadMou96b} that the $\lambda_i$ must satisfy the
consistency equations
$$
2 \lambda_l \lambda_m P^{lm}{}_{jk} - 
\lambda_i F^i{}_{jk} - K_{jk} = 0                                 \eqno(2.19)
$$
where the $F^i{}_{jk}$ are complex numbers.
The structure elements $C^i{}_{jk}$ are defined by the equation
$$
d\theta^i = 
- {1\over 2} C^i{}_{jk} \theta^j \theta^k.                        \eqno(2.20)
$$
From general arguments~\cite{DimMad96, MadMou96b} it follows that
$$
C^i{}_{jk} = F^i{}_{jk} - 2 \lambda_l P^{(li)}{}_{jk}.             \eqno(2.21)
$$
The structure elements are not therefore in general complex numbers.

The simplest noncommutative algebras are the algebras $M_n$ of
$n \times n$ complex matrices.  Let $\lambda_i$ in $M_n$, for 
$1 \leq i \leq n^2-1$, be an antihermitian basis of the Lie algebra of
the special unitary group $SU_n$. The product $\lambda_i \lambda_j$ can
be written in the form
$$
\lambda_i \lambda_j = {1\over 2} C^k{}_{ij} \lambda_k +
{1\over 2} D^k{}_{ij} \lambda_k - {1 \over n} g_{ij}.
$$
The $g_{ij}$ are the components of the Killing metric; we shall use it
to raise and lower indices.  The $C^k{}_{ij}$ here are the structure
constants of the group $SU_n$ and $g_{kl}D^l{}_{ij}$ is trace-free and
symmetric in all pairs of indices. For each $\lambda_i$ 
we introduce derivations $e_i$ as above in Equation~(2.8). In this case
the $e_i$ span the vector space of all derivations~\cite{Dub88} 
of the algebra and form a Lie algebra with commutation relations
$$
[e_i, e_j] = C^k{}_{ij} e_k.
$$
It is an elementary fact of algebra that any derivation $X$ of $M_n$ can
be written as a linear combination $X = X^i e_i$ of the $e_i$ with the
$X^i$ complex numbers. We have now
$$
d\lambda^i(e_j) = [\lambda_j, \lambda^i ] = - C^i{}_{jk}\lambda^k.
$$
The frame~\cite{DubKerMad89} is given by
$$
\theta^i = \lambda_j \lambda^i d\lambda^j.
$$
The corresponding $P^{ij}{}_{kl}$ is given by (2.15) and $\theta$
satisfies
$$
d \theta + \theta^2 = 0.                                          \eqno(2.22)
$$
This is a particular case of (2.18) with $K_{ij} = 0$. We have seen that
as a bimodule $\Omega^1(M_n)$ is generated by $\theta$ alone. For
dimensional reasons $\Omega^1(M_n)$ cannot be of rank one.  In fact the
free $M_n$-bimodule of rank one is of dimension $n^4$ and the dimension
of $\Omega^1(M_n)$ is equal to $(n^2-1)n^2 < n^4$. With the normalization
which we have used for the generators $\lambda_i$ the element
$$
\zeta = {1\over n^2} 1 \otimes 1 - {1\over n} \lambda_i \otimes \lambda^i 
$$
is a projector in $M_n \otimes M_n$ which commutes with the elements of
$M_n$. This can be written~\cite{DubMadMasMou96} as $d(M_n) \zeta = 0$.
We have the direct-sum decomposition
$$
M_n \otimes M_n = \Omega^1(M_n) \oplus M_n \,\zeta.
$$ 

One can use matrix algebras to construct examples of differential
calculi which have nothing to do with derivations.  Consider the algebra
$M_n$ graded as in supersymmetry with even and odd elements
and introduce a graded commutator between two matrices $\alpha$ and
$\beta$ as
$$
[\alpha, \beta] = \alpha \beta - 
(-1)^{\vert \alpha \vert \vert \beta \vert} \beta \alpha
$$
where $\vert \alpha \vert$ is equal to 0 or 1 depending on whether
$\alpha$ is even or odd. One can define on $M_n$ a graded derivation
$\hat d$ by the formula
$$
\hat d \alpha = - [\eta , \alpha],                                \eqno(2.23)
$$
where $\eta$ is an arbitrary antihermitian odd element. Since $\eta$
anti-commutes with itself we find that $\hat d\eta = -2\eta^2$ and for
any $\alpha$ in $M_n$
$$
\hat d^2 \alpha = [\eta^2, \alpha].                              \eqno(2.24)
$$
The grading can be expressed as the direct sum 
$M_n = M_n^+ \oplus M_n^-$ of the even and odd elements of $M_n$.  This
decomposition is the analogue of (2.4).  If $n$ is even it is
possible to impose the condition
$$
\eta^2 = - 1.                                                     \eqno(2.25)
$$
From (2.24) we see that $\hat d^2 = 0$ and $\hat d$ is a differential. In
this case we shall write $\hat d = d$.  We see that $\eta$ must satisfy
$$
d\eta + \eta^2 = 1,                                               \eqno(2.26)
$$
an equation which is to be compared with (2.18) and (2.22).  If we 
define for all $p \geq 0$
$$
\Omega^{2p}(M^+_n) = M^+_n, \qquad 
\Omega^{2p+1}(M^+_n) = M^-_n                                      \eqno(2.27)
$$
then we have defined a differential calculus over $M^+_n$. The 
differential algebra based on derivations can be embedded in a larger
algebra such that a graded extension of (2.16) exists for 
all elements~\cite{Mad95}. In fact any differential calculus 
can be so extended.

As an example let $n=2$. To within a normalization the matrices
$\lambda_i$ can be chosen to be the Pauli matrices. We define
$\lambda_1$ and $\lambda_2$ to be odd and $\lambda_3$ and the identity
even. The most general possible form for $\eta$ is a linear combination
of $\lambda_1$ and $\lambda_2$ and it can be normalized so that
(2.25) is satisfied.  Using $\Omega^*(M_2^+)$ one can construct a
differential calculus over the algebra of functions on a double-sheeted
space-time~\cite{ConLot90, Coq89}. This doubled-sheeted structure permits
one to introduce a description of parity breaking in the weak
interactions.

If $n$ is not even or, in general, if $\eta^2$ is not proportional to
the unit element of $M_n$ then $\hat d^2$ given by (2.24) will
not vanish and $M_n$ will not be a differential algebra.  It is still
possible however to construct over $M_n^+$ a differential calculus
$\Omega^*(M_n^+)$ based on (2.23). Essentially what one does is just
eliminate the elements which are the image of $\hat
d^2$~\cite{ConLot92}. The problem here is an analog of the problem we
mentioned in the commutative case where the square of the Dirac operator
is not proportional to the identity.

As an example let $n=3$. There is a grading defined by the
decomposition $3 = 2 + 1$ The most general possible form for $\eta$ is
$$
\eta = \left(
\begin{array}{ccc}
   0   &    0    & a_1 \\ 
   0   &    0    & a_2 \\ 
-a^*_1 & -a^*_2  &  0
\end{array}\right).                                               \eqno(2.28)
$$
For no values of the $a_i$ is it possible to impose the condition
(2.25). The general construction yields 
$\Omega^0(M_3^+) = M_3^+ = M_2 \times M_1$ and $\Omega^1(M_3^+) = M_3^-$
as in the previous example but after that the elimination of elements
which are the image of $\hat d^2$ reduces the dimensions. One finds
$\Omega^2(M_3^+) = M_1$ and $\Omega^p(M_3^+) = 0$ for 
$p\geq 3$~\cite{ConLot92, Mad95}.

The noncommutative equivalent of a coordinate transformation of
space-time could reasonably be considered to be an automorphism of the
algebra ${\cal A}_\kbar$; if ${\cal A}_\kbar$ corresponds to the
algebra of smooth functions then the corresponding limit coordinate
transformation would then be considered as smooth. There are however
problems with this identification. It can be seen from the
``fuzzy-sphere'' example to be described in Section~5.3 that the algebra of
morphisms is sometimes to small. In this example one does not obtain as
limit a general coordinate transformation~\cite{Mad97b}. At the same
time the algebra of morphisms is too big since some of them in the
commutative limit change even the topology of the limit
manifold~\cite{MadSae97}. The important question is how the
automorphisms are to be extended to the algebra 
$\Omega^*({\cal A}_\kbar)$. Consider as example the differential 
calculus defined in Section~2.7 by the set of $\lambda_i$ and let 
$\lambda_i \mapsto \lambda^\prime_i = u^{-1} \lambda_i u$ be an inner
automorphism of the algebra. Then it is easy to see that 
$$
d\lambda^\prime_i = - [\theta, \lambda^\prime_i].
$$
A sufficient condition then for $u$ to respect the differential structure
would be 
$$
\theta^\prime = \theta.
$$

\section{Classical gravity}

There are several ways to introduce a gravitational field on ordinary,
commutative space-time.  Our main constraint is that we would like it to
be expressed entirely in terms of the algebra of functions ${\cal A}$
and of a differential calculus $\Omega^*({\cal A})$ over ${\cal A}$
since this is what can be generalized to the noncommutative case.  The
geometry of ordinary smooth spaces was written from the point of view of
the algebra of smooth functions by Koszul~\cite{Kos60} in his lectures
at the Tata Institute.  We shall use the moving frame formalism since it
is most convenient for what we believe to be the correct generalization
to the noncommutative case. A moving frame is a nonsingular set of four
1-forms $\theta^\alpha$. Since we are especially interested in
space-times which are near to Minkowski space we can suppose that the
moving frame can be globally defined, an assumption which is a
topological restriction. The $\theta^\alpha$ are of course dual to a set
$e_\beta$ of derivations of ${\cal A}$: 
$$
\theta^\alpha (e_\beta) = \delta^\alpha_\beta.
$$
Equation~(2.6) is a particular case of this with 
$\theta^\alpha = dx^\alpha$.

Let $g_{\alpha\beta}$ be the standard components of the Minkowski
metric. We define a metric $g$ by the condition that the moving frame be
orthonormal:
$$
g(\theta^\alpha \otimes \theta^\beta) = g^{\alpha\beta}.            \eqno(3.1)
$$
If we write in coordinates 
$\theta^\alpha = \theta^\alpha_\lambda dx^\lambda$ then (3.1) is
equivalent to defining the metric by the line element
$ds^2 = g_{\alpha\beta} \, \theta^\alpha_\mu \, \theta^\beta_\nu \,
dx^\mu \otimes dx^\nu$. We extend the metric to all tensor products by
the requirement of linearity:
$$
f g(\theta^\alpha \otimes \theta^\beta) =
g(f \theta^\alpha \otimes \theta^\beta), \qquad
g(\theta^\alpha \otimes \theta^\beta) f =
g(\theta^\alpha \otimes \theta^\beta f).                            \eqno(3.2)
$$
These linearity conditions are equivalent to a locality condition for the
metric; the length of a vector at a given point depends only on the value of
the metric and the vector field at that point.  We shall return to this
in Section~7. The second rule is here a triviality but in the
noncommutative case this will not be so. A metric can be therefore
considered as a bimodule map
$$
\Omega^1({\cal A}) \otimes_{{\cal A}}  \Omega^1({\cal A})
\buildrel g \over \rightarrow {\cal A}.                             \eqno(3.3)
$$
The structure functions $C^\alpha{}_{\beta\gamma}$ are defined as in the
previous section by the equations analog to (2.20).

A covariant derivative or linear connection can be defined a rule 
which associates to each covariant vector $\xi$ a 2-index covariant
tensor $D\xi$. It can be defined on the basis $dx^\lambda$ by
$$
D (dx^\lambda) = - \Gamma^\lambda_{\mu\nu} dx^\mu \otimes dx^\nu
$$
and extended to an arbitrary 1-form $\xi = \xi_\lambda dx^\lambda$ 
by the Leibniz rule:
$$
D \xi = d\xi_\lambda \otimes dx^\lambda + \xi_\lambda D(dx^\lambda)
= d\xi_\lambda \otimes dx^\lambda
- \xi_\lambda \Gamma^\lambda_{\mu\nu} dx^\mu \otimes dx^\nu 
$$
It can also be written in terms of the moving frame. We define the Ricci
rotation coefficients $\omega^\alpha{}_{\beta\gamma}$ by the equation
$$
D \theta^\alpha = - \omega^\alpha{}_{\beta\gamma} 
\theta^\beta \otimes \theta^\gamma.
$$
A covariant derivative can be defined as a map
$$
\Omega^1({\cal A}) \buildrel D \over \rightarrow
\Omega^1({\cal A}) \otimes_{\cal A} \Omega^1({\cal A})             \eqno(3.4)
$$
which satisfies the Leibniz rules
$$
D (f \xi) = df \otimes \xi + f D\xi, \qquad
D (\xi f) = D (f \xi).                                              \eqno(3,5)
$$
The second rule is here a triviality but in the noncommutative case it
will have to be modified.

Using a graded Leibniz rule, $D$ can be extended to higher-order
forms and the curvature 2-form $\Omega^\alpha{}_\beta$ defined by 
the equation
$$
D^2 \xi = - \xi_\alpha \Omega^\alpha{}_\beta \otimes \theta^\beta.
$$
The curvature is the field strength of the gravitational field. The
minus sign is an historical convention. One can be write 
$\Omega^\alpha{}_\beta$ in terms of the basis as
$$
\Omega^\alpha{}_\beta = 
{1\over 2} R^\alpha{}_{\beta\gamma\delta} \theta^\gamma \theta^\delta
$$
an equation which defines the components
$R^\alpha{}_{\beta\gamma\delta}$ of the Riemann tensor.

We have two maps of $\Omega^1({\cal A})$ into $\Omega^1({\cal A})$, the
composite map $\pi \circ D$ as well as the exterior derivative. The
difference between the two is the torsion:
$$
T = d - \pi \circ D.                                                \eqno(3.6)
$$
In particular
$$
T(dx^\lambda) = {1\over 2} \Gamma^\lambda_{[\mu\nu]} dx^\mu dx^\nu.
$$
It vanishes with the antisymmetric part of $\Gamma^\lambda_{\mu\nu}$.

Let $\xi$ and $\eta$ be 1-forms and introduce a flip $\sigma$ in 
the tensor product: $\sigma(\xi \otimes \eta) = \eta \otimes \xi$. Then
the covariant derivative can be extended to arbitrary tensors by the
twisted Leibniz rule.
$$
D(\xi \otimes \eta) = D\xi \otimes \eta + 
(\sigma \otimes 1) (\xi \otimes D\eta).
$$
A covariant derivative is said to be compatible with the metric $g$
if 
$$
(1 \otimes g)(D(\theta^\alpha \otimes \theta^\beta)) \equiv
- \omega^\alpha{}_{\gamma\delta} \, \theta^\gamma \,
g(\theta^\delta \otimes \theta^\beta) - \omega^\beta{}_{\gamma\delta}
\, \theta^\gamma \,
g(\theta^\alpha \otimes \theta^\delta) = 0.                        \eqno(3.7)
$$
This will be the case if and only if
$$
\omega_{\alpha\beta\gamma} + \omega_{\gamma\beta\alpha} = 0.        \eqno(3.8)
$$
A metric-compatible $D$ without torsion is completely 
determined by the structure functions:
$$
\omega^\alpha{}_{\beta\gamma} = {1\over 2}
(C^\alpha{}_{\beta\gamma} - C_{\beta\gamma}{}^\alpha + 
C_\gamma{}^\alpha{}_\beta).                                         \eqno(3.9)
$$

The classical theory of gravity involves also a set field equations for
the metric, which are normally supposed to be derived from an action
principle. We shall return to this in Section~7.

\section{Noncommutative gravity}

In formulating a noncommutative theory of gravity we shall be as
conservative as possible and change the definitions of the previous
section only where absolutely necessary. Instead of the commutative 
algebra ${\cal A}$ we have now the noncommutative algebra 
${\cal A}_\kbar$. We have also {\it ipso facto} the Kaluza-Klein 
extension $V_0$ of space-time. Let $e_i$ be a set of derivations and
$\Omega^*({\cal A}_\kbar)$ the corresponding differential calculus as
defined in Section~2. The frame $\theta^i$ plays then the role of the
moving frame in the commutative case. One might think of $n$ as the
`dimension' but this is a delicate issue. One can only say that 
$$
n \geq \hbox{dim} V_0.
$$
Let $g_{ij}$ be the standard components of the Minkowski
metric on a $d$-dimensional extension of space-time. We define again a 
metric $g$ by the condition that the moving frame be
orthonormal:
$$
g(\theta^i \otimes \theta^j) = g^{ij}.                             \eqno(4.1)
$$
We require as before that
$$
f g(\theta^i \otimes \theta^j) =
g(f \theta^i \otimes \theta^j), \qquad
g(\theta^i \otimes \theta^j) f =
g(\theta^i \otimes \theta^j f)                                     \eqno(4.2)
$$
and therefore a metric can be defined as a bimodule map
$$
\Omega^1({\cal A}_\kbar) \otimes_{{\cal A}_\kbar}  \Omega^1({\cal A}_\kbar)
\buildrel g \over \rightarrow {\cal A}_\kbar.                      \eqno(4.3)
$$
Because of the bilinearity and because of the relation (2.10) the
coefficients $g^{ij}$ are here necessarily real numbers. In the
commutative case they could have been chosen as arbitrary functions
and using this freedom one can construct an arbitrary metric 
$g^\prime$ using the moving frame $\theta^\alpha$.  This is a very 
important difference between the commutative and the noncommutative
case. It is the reason why there is essentially only one metric
associated to each differential calculus.  The structure elements
$C^i{}_{jk}$ are given by (2.21). It follows that they will be 
necessarily real numbers if the elements of the frame anticommute.

Let $\sigma$ be an ${\cal A}_\kbar$-bilinear map
$$
\Omega^1({\cal A}_\kbar) \otimes_{{\cal A}_\kbar} \Omega^1({\cal A}_\kbar)
\buildrel \sigma \over \longrightarrow 
\Omega^1({\cal A}_\kbar) \otimes_{{\cal A}_\kbar} \Omega^1({\cal A}_\kbar).
                                                                    \eqno(4.4)
$$
We shall define a linear connection as a covariant derivative $D$
$$
\Omega^1({\cal A}_\kbar) \buildrel D \over \rightarrow
\Omega^1({\cal A}_\kbar) \otimes_{{\cal A}_\kbar} \Omega^1({\cal A}_\kbar)
                                                                    \eqno(4.5)
$$
and a map $\sigma$ such that the following Leibniz
rules~\cite{DubMadMasMou95, Mou95, DubMadMasMou96} are satisfied:
$$
D (f \xi) = df \otimes \xi + f D\xi, \qquad
D (\xi f) = \sigma (\xi \otimes df) + (D\xi) f.                     \eqno(4.6)
$$

The purpose of the $\sigma$ in the second equation is to place the
differential in the first term to the left where it belongs while
respecting the order of the various terms. A description of the bimodule
structure of the module of 1-forms can be given in terms of a
left-module structure with respect to a larger algebra and this leads to
a natural splitting of the covariant derivative $D$ as the sum of two
terms~\cite{DubMadMasMou96}. In the commutative case it is easy to see
that necessarily $\sigma$ is the flip of the previous section.  As in
the commutative case we introduce the elements $\omega^i{}_{jk}$ by the
equation
$$
D \theta^i = - \omega^i{}_{jk} \theta^j \otimes \theta^k.           \eqno(4.7)
$$

There is no consensus at the moment concerning the necessity of two
Leibniz rules. There are authors who maintain~\cite{ChaFelFro93,
LanNguWal94, Sit94, ChaFroGra95, FroGraRec97, HecSch97} that it suffices
to require that the covariant derivative satisfy a left (or right)
Leibniz rule.  There are others who propose~\cite{CunQui95, DabHajLanSin96,
Haj96} introducing both a left and right covariant derivative depending
on which (left or right) Leibniz rule one chooses to enforce.  There are
interesting noncommutative cases~\cite{Mad89b} where the two Leibniz
rules are equivalent, We maintain~\cite{Mou95, MadMasMou95,
DubMadMasMou95, GeoMadMasMou97, DubMadMasMou96, DimMad96} that without
both rules it is not possible to correctly impose a reality condition on
the linear connection~\cite{KasMadTes97} nor will it be possible to
construct nontrivial invariants to serve, for example as lagrangians. We
shall discuss this second point in Section~7.

The torsion is defined exactly as in the commutative case (3.6).
It is straightforward~\cite{DubMadMasMou95, Mou95, DubMadMasMou96} 
to see that if the torsion is to be a bilinear map then the $\sigma$
must satisfy the condition
$$
\pi \circ (1 + \sigma) = 0.                                        \eqno(4.8)
$$
This condition is trivially satisfied in the commutative case. The most
general such $\sigma$ is of the form~\cite{MadMou96b}
$$
\sigma = (1 - \pi) \circ \tau - 1
$$
where $\tau$ is an arbitrary ${\cal A}_\kbar$-bilinear map of
$\Omega^1({\cal A}_\kbar) \otimes_{{\cal A}_\kbar} \Omega^1({\cal A}_\kbar)$
into itself. If $\tau = 2$ then $\sigma^2 = 1$. The condition
that the connection be metric-compatible can be formulated exactly as in
the commutative case:
$$
(1 \otimes g) D (\theta^i \otimes \theta^j) = 0.
$$

If we define the complex numbers $S^{ij}{}_{kl}$ by the equation
$$
\sigma (\theta^i \otimes \theta^j))
= 
S^{ij}{}_{kl} \theta^k \otimes \theta^l                            \eqno(4.9)
$$
then the condition of metric compatibility becomes~\cite{MadMou96b}
$$
\omega^i{}_{jk} + \omega_{kl}{}^m S^{il}{}_{jm} = 0.
$$
This is analogous to the condition (3.8) of the commutative case but
twisted by $\sigma$. We shall discuss the noncommutative generalization
of curvature in Section~7.

For each differential calculus there is a linear connection defined in
term of the form $\theta$~\cite{DubMadMasMou96} given by
$$
D\theta^i = - \theta \otimes \theta^i + \sigma(\theta^i \otimes \theta).
                                                                  \eqno(4.10)
$$
One verifies immediately that it satisfies the two Leibniz rules. In
general it is neither torsion-free nor compatible with the metric.

Suppose that the algebra ${\cal A}_\kbar$ is such that
$[q^\lambda, q^{\mu\nu}] = 0$ and that the matrix $q^{\mu\nu}$ has an
inverse $q^{-1}_{\mu\nu}$. Define the differential structure by choosing
$n=4$ and 
$$
\lambda_\mu = {1\over i\kbar} q^{-1}_{\mu\nu} q^\nu.
$$
It follows that the frame is given by
$$
\theta^\mu = dq^\mu.
$$
The unique torsion-free linear connection compatible with the
corresponding metric is the trivial connection given by~\cite{MadMou96b}
$D\theta^\mu = 0$. Apart from this example there are no non-trivial
examples of linear connections on differential calculi over algebras
which have anything to do with space-time. For this reason we must in
the following section consider simpler models. It is hoped that there
will eventually be a relation between noncommutative gravity and the
quantum theory of gravity, whatever that may be. Speculations have been
made~\cite{MadMou95, Bal97} along these lines.

\section{Models}

\subsection{Lattice models}

One of the advantages of noncommutative geometry is that it gives a
prescription of how one can construct differential calculi over discrete
structures, a construction which involves essentially using the universal
calculus or some quotient of it over the algebra of functions on a
finite set of points. This has been studied from a variety of points of
view~\cite{DimMul93, DimMul94, BimLizSpa94, BalBimLanLizTeo96,
BimErcLanLizSpa96} and recently a book~\cite{Lan97} has appeared to
which we refer for further details. A comparison has yet to be made with
the classical Regge calculus. One of the reasons for this is the
difficulty in defining linear curvature within the context of
noncommutative geometry. We shall return to this problem in Section~7.

\subsection{$q$-models}

The quantum plane is the algebra ${\cal A}$ generated by `variables' 
$x$ and $y$ which satisfy the relation
$$
xy = qyx,                                                       \eqno(5.1)
$$
where $q$ is an arbitrary complex number. As usual it has over it many
differential calculi $\Omega^*({\cal A})$. The commutation relations in
$\Omega^1({\cal A})$ must be consistent with (5.1) but this
condition is not enough to uniquely define the calculus. There is
however a particularly interesting calculus known as the Wess-Zumino
calculus~\cite{PusWor89, WesZum90} which is covariant under the co-action
of a quantum group~\cite{Wor87}. Since the elements of ${\cal A}$ do not
in general commute the elements of $\Omega^1({\cal A})$ will not in
general anti-commute. It has been shown~\cite{DubMadMasMou95} that
consistent with the Wess-Zumino calculus there is a 1-parameter family
of linear connections which is without torsion but not compatible with a
metric. The earliest example of a classical field theory on a
noncommutative structure was furnished~\cite{ConRie87, Con88} by the
electromagnetic field on a particular version of the quantum plane known
as the noncommutative torus. In this case a careful analysis of the
problem posed by the definition of the action was made.  We shall return
to this problem in Section~7.

One can extend the previous algebra by adding the inverses $x^{-1}$ and
$y^{-1}$. For each integer $n$ and each set of $n$ linear-independent
elements $\lambda_i$ of ${\cal A}$, there exists then a differential
calculus $\Omega^*({\cal A})$ based on the derivations 
$e_i f = [\lambda_i, f]$ as in Section~2.2.  Unless however $n = 2$ the
frame has a singular limit as $q \rightarrow 1$. For $n=2$ and a special
choice of the elements $\lambda_i$, given for $q^4 \neq 1$ by
$$
\lambda_1 = {1 \over q^4 - 1} x^{-2} y^2,\qquad
\lambda_2 = {1 \over q^4 - 1} x^{-2},                            \eqno(5.2)
$$
the resulting differential calculus is an extension of the Wess-Zumino
calculus.  The normalization has been chosen so that the structure
elements $C^i{}_{jk}$ contain no factors $q$. The corresponding frame
(2.9) is given by
$$
\theta^1 = - q^4 (q^2 + 1) x y^{-2} dx,\qquad
\theta^2 = - q^2 (q^2 + 1) x (x y^{-1} dy - dx).                  \eqno(5.3)
$$
It satisfies the commutation relations
$$
(\theta^1)^2=0,\qquad (\theta^2)^2=0, \qquad
q^4 \theta^1\theta^2 + \theta^2\theta^1 = 0.
$$
These relations determine the structure of the algebra 
$\Omega^*({\cal A})$.  The corresponding $\theta$ cannot be considered
as an element of the Wess-Zumino calculus since the $\theta^i$ are
constructed using the inverses of $x$ and $y$. The $\lambda_i$ satisfy
an equation of the form (2.19) with $F^i{}_{jk} = 0$, $K_{ij} = 0$
and $P^{lm}{}_{jk}$ defined in terms of the $R$-matrix of the associated
quantum group.

Consider the quantum groups $GL_q(n)$ with generators $T^i_j$
and antipode $\kappa$.  The left-invariant 1-forms
$$
\omega^i_j = \kappa(T^i_k) dT^k_j
$$
generate~\cite{Wor87} a bicovariant differential calculus
$\Omega^*(GL_q(n))$.  The exterior derivative is defined with the help
of the right and left-invariant 1-form $\theta$.  If $\sigma$ is any
generalized permutation then the map
$$
\nabla^\sigma: \Omega^1(GL_q(n)) \rightarrow 
\Omega^1(GL_q(n)) \otimes_{\cal A} \Omega^1(GL_q(n))
$$
defined by
$$
\nabla^\sigma(\omega) = - \theta \otimes \omega + 
\sigma(\omega \otimes \theta)                                     \eqno(5.4)
$$
defines a linear connection~\cite{GeoMadMasMou97} associated to
$\sigma$.  This is to be compared with (4.10). It can be shown that for
each $\sigma$, the only linear connection for generic $q$ is the one
defined by (5.4).  Further it can be shown that it has necessarily
vanishing torsion.  This is in contrast to the commutative case where
there are an infinite number of linear connections not necessarily
bicovariant nor torsion-free and where the generalized permutation is
constrained to be the ordinary permutation.  It is also in contrast to
the cases with $q$ a root of unity.  The arbitrariness in the deformed
case lies merely in the generalized permutation for which it can be
shown that there is at least a 2-parameter family, functions of $q$.
The commutative limit is non-singular for a class of such functions
which tend to the identity when $q \rightarrow 1$.  More details are to
be found in the article by Georgelin {\it et al.}~\cite{GeoMadMasMou97}.
See also the article by Heckenberger \& Schm\"udgen~\cite{HecSch97}.

For more details of $q$-deformed spaces and the groups which act on them
we refer to the Carg\`ese lectures of Zumino~\cite{ChuHoZum96}.  For a
discussion of the possible $q$ deformations of Minkowski space we refer
to the literature~\cite{FicLorWes95, KehMeeZou95, AscCas96, Pod96,
PodWor96}.  These $q$ deformations has been considered~\cite{Maj97} as
possible regulators of ultraviolet divergences in the sense of Snyder
but as yet no linear connections have been constructed over them.  The
differential geometry of the $h$-deformed quantum plane~\cite{Kup92,
Agh93} has been recently studied~\cite{ChoMadPar97}; it is a
noncommutative version of the Poincar\'e half-plane.

\subsection{Finite models}

Linear connections have been constructed~\cite{MadMasMou95} over the
finite differential calculi defined at the end of Section~2.2.  The
calculi based on derivations can be shown to have many linear
connections but only one which is torsion-free and metric compatible.
For $n=2$ the calculus (2.27) is the universal calculus over the algebra
$M^+_2$ and it admits only the trivial connection. This model has been
shown~\cite{MadMouSit97} to be a singular contraction of the
$n=2$ model based on derivations. The example with $n=3$ and
differential calculus defined by (2.28) possesses a 1-parameter family
of generalized permutations $\sigma$ (4.4) and for each of these there
is a unique covariant derivative (4.5). All of these linear connections
are torsion-free. For a special value of the parameter of $\sigma$ the
connection is compatible with a metric.

The $SO_3$-invariant `lattice structure' referred to in the Introduction
can be constructed using the formalism of Section~2.2. For this we let 
$\lambda_i$ be a set of three antihermitian generators of the
irreducible $n$-dimensional representation of the Lie algebra of $SU_2$. 
To discuss the commutative limit it is convenient to change the
normalization of the generators $\lambda_i$. We introduce the parameter
$\kbar$ with the dimensions of $(\hbox{length})^2$ and define 
`coordinates' $x_i$ by
$$
x_i = i \kbar \lambda_i.
$$
The $x_i$ satisfy therefore the commutation relations
$$
[x_i , x_j ] = i \kbar x_k \, C^k{}_{ij}.                          \eqno(5.5)
$$
We choose the $\lambda_i$ so that $C_{ijk} = r^{-1} \epsilon_{ijk}$
where $r$ is a length parameter. These structure constants are in
general independent from the structure elements defined by (2.21) but in
the present case they are equal.  Introduce the $SU_2$-Casimir metric
$g_{ij}$. The matrix $g^{ij} x_i x_j$ is the Casimir operator. We choose
$\kbar$ so that $g^{ij} x_i x_j = r^2$. We have then from (5.5) the
relation
$$
4 r^4 = (n^2-1) \kbar^2.                                           \eqno(5.6)
$$
The commutative limit is the limit $\kbar \rightarrow 0$. Were we
considering a noncommutative model of space-time then we would be
tempted to identify $\kbar$ with the inverse of the square of the Planck
mass, $\kbar = \mu^{-2}_P$, and consider space-time as fundamentally
noncommutative in the presence of gravity.

The differential calculus has a basis~\cite{Mad92a, Mad92b}

$$
\theta^i = - C^i{}_{jk} x^j dx^k - i\kbar r^{-2} \theta x^i.       \eqno(5.7)
$$
The 1-form $\theta$ can be written
$$
\theta = i \kbar^{-1} x_i \theta^i = r^2 \kbar^{-2} x_i dx^i.      \eqno(5.8)
$$
In the commutative limit $\theta$ diverges but $\kbar \theta \rightarrow
r^2 A$ where $A$ is the Dirac-monopole potential of unit magnetic
charge.  The commutative limit of the frame $\theta^i$ is a moving frame
on a $U_1$-bundle over $S^2$. In this case it is not the frame bundle. A
standard Kaluza-Klein reduction gives rise to the potential $A$ as well
as the geometry of the sphere. We refer to this structure as the `fuzzy
sphere'. Various field theories have been studied~\cite{GroMad92, 
GroPre95} on it and it has been generalized~\cite{GroPre93, GroKliPre96, 
GroKliPre97a, GroKliPre97c, CarWat97} in several ways.  

The linear connection on the fuzzy sphere is the same~\cite{MadMasMou95}
as that of the sphere.  Recently over the same matrix algebra but with
another differential calculus, obtained by using another solution to
Equation~(2.19) a less trivial connection has been
constructed~\cite{Mad97b} whose curvature is not invariant under the
action of the rotation group.

Although we are primarily interested in the matrix version of surfaces
as an model of an eventual noncommutative theory of gravity they have a
certain interest in other, closely related, domains of physics. Without
the differential calculus the fuzzy sphere is basically just an
approximation to a classical spin $r$ by a quantum spin $r$ given by
(5.6) with $\hbar$ in lieu of $\kbar$.  It has been extended in various
directions under various names and for various reasons~\cite{Ber75,
deWHopNic88, Hop89, FaiFletZac89, CahGutRaw90, BorHopSchSch91}.  In
order to explain the finite entropy of a black hole it has been
conjectured, for example by 't~Hooft~\cite{'tH96}, that the horizon has
a structure of a fuzzy 2-sphere since the latter has a finite number of
`points' and yet has an $SO_3$-invariant geometry.  The horizon of a
black hole might be a unique situation in which one can actually `see'
the cellular structure of space. Matrices can also be used to give a
finite `fuzzy' description of the space complementary to a Dirichlet
$p$-brane, a description which will allow one perhaps to include the
reasonable property that points should be intrinsically `fuzzy' at the
Planck scale.  This has much in common with the noncommutative version
of Kaluza-Klein theory which we shall describe in the next section.
Strings naturally play a special role here since they have a world
surface of dimension two and an arbitrary matrix can always be written
as a polynomial in two given matrices. We refer to the literature for a
description of Dirichlet branes in general~\cite{Pol96, BonChu97, Dij97}
and within the context of $M$(atrix)-theory~\cite{BanFisSheSus96,
GanRamTay96, HoWu96, Ban97}.  The action of the matrix description of
the complementary space is conjectured~\cite{deWHopNic88} to be
associated to the action in the infinite-momentum frame of a
super-membrane of dimension $p$. Since quite generally the compactified
factors of the surfaces normal to the $p$-branes are of the Planck scale
we conclude~\cite{MadSae97} that they have ill-defined topology and that
a matrix description will include a sum over many topologies.  Attempts
have been made to endow them with a smooth differential
structure~\cite{Mad96, GroKliPre97b}.  Speculations have also been
made~\cite{AreVol97} concerning their relation with knots.

We have already mentioned that several models for the algebraic structure
of space-time have been proposed~\cite{Sny47a, Mad89a, DopFreRob95,
DubKerMad97} but there have been few discussions~\cite{MadMou96b, Mad97a}
of associated differential structures and at present no interesting
examples~\cite{MadMou96b} of linear connections.

\section{Kaluza-Klein theory}

Although the ultimate ambition of noncommutative geometry (in physics)
is is to introduce a noncommutative version of space-time and to use it
to describe quantum gravity, one can consider the much more modest task
of introducing a modified version of Kaluza-Klein theory in which the
hidden `internal' space alone is described by a noncommutative geometry.
In traditional Kaluza-Klein theory~\cite{App87, BaiLov87, CoqJad88} the
higher-order modes in the mode expansion of the field variables in the
coordinates of the internal space are neglected, with the justification
that they have all masses of the order of the Planck mass and would not
be of interest in conventional physics. The alternative theory we here
propose possesses {\it ab initio} only a finite number of modes; there
are no extraneous modes to truncate.  We would like to suggest also that
the noncommutative version of Kaluza-Klein theory is more natural than
the traditional one in that a hand-waving argument~\cite{MadMou95} can
be given which allows one to think of the extra algebraic structure as
being due to quantum fluctuations of the light-cone in ordinary
4-dimensional space-time. We already suggested in the Introduction that
this might be the origin of the noncommutative structure of space-time
itself.

We suppose then that the algebra ${\cal A}_\kbar$ has the structure of a
tensor product
$$
{\cal A}_\kbar = {\cal C}(V) \otimes M_n                       \eqno(6.1)
$$
of an algebra of smooth functions on space-time $V$ and a matrix algebra
$M_n$. We introduce a differential calculus $\Omega^*({\cal A}_\kbar)$
over ${\cal A}_\kbar$ which is a tensor product of the de~Rham
differential calculus $\Omega^*(V)$ over $V$ and a differential calculus
$\Omega^*(M_n)$ over the matrix factor.  If we define
$$
\Omega^1_h = \Omega^1(V) \otimes M_n,  \qquad
\Omega^1_v = {\cal C}(V) \otimes \Omega^1(M_n),
$$
we can write $\Omega^1({\cal A})$ as a direct sum: 
$$
\Omega^1({\cal A}) = \Omega^1_h \oplus \Omega^1_v.             \eqno(6.2)
$$
of two terms, the horizontal and vertical
parts, using notation from traditional Kaluza-Klein theory.
The exterior derivative $df$ of an element $f$ of ${\cal A}_\kbar$ 
has a similar decomposition
$$
df = d_hf + d_vf.                                              \eqno(6.3)
$$
We can choose $\theta^i = (\theta^\alpha, \theta^a)$ as a basis for
$\Omega^1({\cal A})$ where $\theta^\alpha$ is a moving frame on $V$,
supposed for convenience to be parallelizable and $\theta^a$ is a frame
of the sort introduced in Section~2.2

Using the above differential structure one can study electromagnetism as
well as gravity. We consider first the former.  Most of the efforts to
introduce noncommutative geometry into particle physics have been in
fact directed towards trying to find an appropriate noncommutative
generalization of an old idea~\cite{ForMan80, Man79, ChaMan80, Fai79} to
try to unify Yang-Mills and Higgs fields by studying electromagnetism in
higher dimensions.  We write the electromagnetic field strength $F$ as
$$
F = {1 \over 2} F_{ij} \theta^i \theta^j.
$$ 
Then the electromagnetic action on ${\cal A}_\kbar$ takes the form
$$
S = {1 \over 4} \int_V \tr (F_{ij} F^{ij}).                     \eqno(6.4)
$$
The trace over the matrix factor is the equivalent of the integral over
space-time. 

Let $\omega$ be the electromagnetic potential, which we
decompose $\omega = \omega_h + \omega_v$ in typical Kaluza-Klein fashion
as the sum of a horizontal component and a vertical component. 
The gauge transformations are the unitary elements ${\cal U}_n$ of
${\cal A}_\kbar$. We notice that the form $\theta$ which we introduced
in Section~2.2 is gauge-invariant~\cite{DubKerMad89}. It is natural then
to decompose $\omega_v$ as a sum $\omega_v = \theta + \phi$ where $\phi$
is the difference between two gauge potentials and so transforms
covariantly under a gauge transformation. After a short calculation one
arrives~\cite{DubKerMad89} at a unification of Yang-Mills and Higgs
fields with the potential of the Higgs particle given by the curvature
of the covariant derivative in the algebraic `directions'. One
calculates how the particle and mass spectra vary as one varies the
extra noncommutative algebra and the associated differential calculi.
Much ingenuity has gone into these calculations which often involve very
sophisticated mathematics but which ultimately reduce to simple
manipulations with matrices.

The simplest and most intuitive models are those which use differential
calculi based on derivations~\cite{DubKerMad89, BalGurWal91}, More
general calculi constructed directly from a Dirac operator without the
use of derivations, are less rigid and can be chosen so that the
resulting action coincides with that of the Standard Model. The first
example~\cite{ConLot90, Coq89} was based on the differential calculus
defined by Equation~(2.23) for $n=2$.  The extension~\cite{ConLot92} to
$n=3$ and higher~\cite{CoqHauSch95, LizManMieSpa1996, PriSch97} soon
followed. There exist several reviews~\cite{Kas93, VarGra93} of these
models. A comparison of the two approaches has been
given~\cite{MadMouSit97} in a simple case.  The weak interactions
violate parity and this fact must be included in a realistic model. No
derivation-based model with explicit parity violation has been
developed; the models mentioned above rely implicitly on spontaneous
parity-breaking mechanisms like the `see-saw' mechanism.  The
double-sheeted structure of the Dirac-based models lends itself more
readily to the introduction of explicit parity violation.

Very few of the results of the preceding subsection can be developed
within the context of the theory of gravity and none of them have as yet
any significance for particle physics. We refer simply to the original
literature.  Gravity was included~\cite{Mad89b} in the first
noncommutative version of Kaluza-Klein theory and developed~\cite{Mad90,
MadMou93} in subsequent articles.  Recent reviews~\cite{Mad95, MadMou96a,
FroGraRec97} are to be found. We have already mentioned in Section~4
that there is no consensus concerning the definition of a linear
connection and we mention in the next section the problems concerning
the definition of curvature and the choice of action functional.

\section{Open Problems}

The fundamental open problem of the noncommutative theory of gravity
concerns of course the relation it might have to a future quantum theory
of gravity either directly or via the theory of strings and membranes as
mentioned at the end of Section~6. But there are more immediate
technical problems which have not received a satisfactory answer. The
most important ones concern the definition of the curvature. It is not
certain that the ordinary definition of curvature taken directly from
differential geometry is the quantity which is most useful in the
noncommutative theory. The main interest of curvature in the case of a
smooth manifold definition of space-time is the fact that it is local.
Riemann curvature can be defined as a map
$$
\Omega^1({\cal C}(V)) {\buildrel R \over \longrightarrow} 
\Omega^2({\cal C}(V)) \otimes_{{\cal C}(V)} \Omega^1({\cal C}(V)).
$$
If $\xi \in \Omega^1({\cal C}(V))$ then $R(\xi)$ at a given
point it depends only on the value of $\xi$ at that point. This can be
expressed as a bilinearity condition; the above map is a 
${\cal C}(V)$-bimodule map.  If $f \in {\cal C}(V)$ then
$$
f R(\xi) = R(f\xi), \qquad  R(\xi f) = R(\xi) f.                   \eqno(7.1)
$$
In the noncommutative case bilinearity is the natural (and only
possible) expression of locality. It has not yet been possible to
enforce it in a satisfactory manner~\cite{DubMadMasMou96}.

In the noncommutative case considered in Section~2.2, where the module of
1-forms is free one, can formally identify the curvature as usual with
the operator $D^2$ and set
$$
D^2 \theta^i = - {1 \over 2} R^i{}_{jkl}
\theta^k \theta^l \otimes \theta^j.                                \eqno(7.2)
$$
Since $D^2$ is not necessarily right-linear as an operator we cannot
conclude that the coefficients $R^a{}_{bcd}$ necessarily lie in the
center of the algebra.

We define the Ricci map
$$
\Omega^1 \buildrel  \mbox{Ric} \over \longrightarrow   \Omega^1
$$
by $\mbox{Ric} = - (1\otimes g) \circ D^2$. In terms of the frame we have
$$
\mbox{Ric} \, (\theta^i) = 
{1 \over 2} R^i{}_{jkl} \theta^k g(\theta^l \otimes \theta^j).
$$
It is given by 
$$
\mbox{Ric} \, (\theta^i) = R^i{}_j \theta^k.                     \eqno(7.3)
$$
Formally then one can write vacuum field equations as 
$$
\mbox{Ric} = 0.
$$

We are unable at the moment to propose a satisfactory definition of an
action which would yield as field equations the vanishing of the Ricci
map. and indeed we are not in a position to argue that there is even a
valid action principle. A discussion of this point has been made by
Connes and coworkers in a series of articles~\cite{Con88, KalWal95,
AckTol96, Con96, FroGraRec97} based on an idea of Sakarov~\cite{Sak75}
applied to a Kaluza-Klein theory similar to the one described in the
previous section.  The definition which these authors propose is valid
only on the noncommutative generalizations of compact spaces with
euclidean-signature metrics.  Cyclic homology groups have been
proposed~\cite{Con86} as the appropriate generalization to
noncommutative geometry of topological invariants; the appropriate
definition of other, non-topological, invariants in not clear.

\section{Mathematics}

At a more sophisticated level one would have to add a topology to ${\cal
A}_\kbar$ and consider a closed algebra. Since we have identified the
generators as hermitian operators on a Hilbert space, the most obvious
structure would be that of a von~Neumann algebra. We refer to
Connes~\cite{Con94} for a description of these algebras within the
context of noncommutative geometry. A large part of the interest of
mathematicians in noncommutative geometry has been concerned with the
generalization of topological invariants~\cite{Con86, CunQui95, Mos97}
to the noncommutative case. It was indeed this which lead Connes to
introduce and develop cyclic cohomology. Another interest has been the
generalization of the idea introduced by Atiyah of an homology theory
dual to the $K$-theory of vector bundles. The fundamental object here is
a $K$-cycle or spectral triple, a set $({\cal A}, D, {\cal H})$
consisting of an associative algebra ${\cal A}$ with a representation on
a Hilbert space ${\cal H}$ and a `Dirac operator' $D$ to define a
differential calculus. All the examples of differential calculi which we
have considered here can be formulated as spectral triplets; the
simplest was given in Section~2.2 with ${\cal A} = M_2^+$, $D = \eta$
and ${\cal H} = {\mathbb C}^2$.  Connes~\cite{ConRie87, Con88} has also
developed and extended the notion of a Dixmier trace on certain types of
algebras as a possible generalization of the notion of an integral.  The
mathematics of quantum groups and quantum spaces has also been
considerably studied. We refer, for example, to the book by
Majid~\cite{Maj95}.

\section*{Acknowledgments} 

Part of this research was done while the author was visiting the Erwin
Schr\"odinger Institute, Vienna. The author would like to thank W.
Thirring for his hospitality. He would also like to thank G. Goldin and
D. Lambert for interesting conversations.

\end{document}